\documentclass[prl,preprint,showpacs,showkeys]{revtex4}
\usepackage{graphicx}
\begin{document}

\preprint{nucl-th/0506073}

\title{Suppressed fusion cross section for neutron halo nuclei}

\author{M. Ito\footnote{Corresponding author,
             Email: itom@nucl.ph.tsukuba.ac.jp;
             TEL: +81 29-853-4257;
             FAX: +81 29-853-4492}}
\affiliation{Institute of Physics, University of Tsukuba, 
             Tsukuba 305-8571, Japan}
\author{K. Yabana, T. Nakatsukasa}
\affiliation{Institute of Physics
  and Center for Computational Sciences, University of Tsukuba, 
  Tsukuba 305-8571, Japan}
\author{M. Ueda}
 \affiliation{Akita National College of Technology, Akita 011-8511, Japan}

\date{\today}

\begin{abstract}
Fusion reactions of neutron-halo nuclei are investigated theoretically
with a three-body model. The time-dependent wave-packet method is used 
to solve the three-body Schr\"odinger equation.
The halo neutron behaves as a spectator during the
Coulomb dissociation process of the projectile.
The fusion cross sections of $^{11}$Be - $^{209}$Bi and 
$^6$He - $^{238}$U are calculated and are compared with
measurements. Our calculation indicates that the fusion 
cross section is slightly hindered by the presence of 
weakly bound neutrons.
\end{abstract}

\pacs{25.60.Pj, 25.70.Mn,24.10.-i}
\keywords{NUCLEAR REACTION, $^{209}$Bi($^{10}$Be,X), ($^{11}$Be,X),
          $^{238}$U($^4$He,X), ($^6$He,X),
          Calculated fusion cross section, Three-body model,
          Time-dependent wave-packet method}
\maketitle

Physics at the dripline is one of current subjects in nuclear physics.
Adding neutrons to a nucleus as many as possible,
we often find that some neutrons are bound very weakly.
These neutrons form a spatially extended neutron
cloud which is called the neutron halo. The halo nuclei have
been attracting many researchers since its discovery in 
the secondary beam experiment \cite{Tanihata85}. The neutron-halo nucleus
exhibits many properties different from normal nuclei.
For instance, it violates the nuclear
saturation of density and binding energy. 
Experimentally, these nuclei are mostly studied with reactions using
the secondary beam.

Since the halo nuclei are weakly bound
and easily break up, developments of the reaction theories capable 
of describing coupling to the continuum channels are required.
For reactions in medium and high incident energies, the Glauber and 
the eikonal theories have been successful 
\cite{Suzuki03, Bertsch89, Ogawa92, Yabana92}. 
In these theories, the wave function of the halo neutron in the projectile
is frozen during the reaction.
The success of the eikonal theories rests on the fact that
the dynamics can be well separated into fast and slow motions.
Namely, the relative motion of projectile and target is much faster than
that of halo neutrons.
There is, however, no such clear distinction in collisions at low energy.
In spite of many experimental and theoretical efforts in the last decade, 
the reaction mechanism of halo nuclei at low incident energies is still
an open question.

The fusion of halo nuclei shows an example of the controversial subjects.
In early stages, a simple and intuitive theoretical argument has been proposed
\cite{Takigawa91, Hussein92}:
The fusion probability was expected to be enhanced because of
the spatially extended density of the halo neutron. A further enhancement
was predicted by the coupling to soft modes inherent to the halo structure. 
Breakup effects were predicted to be small,
sustaining a large enhancement of fusion cross section especially at
sub-barrier energies \cite{Hussein92,Takigawa93}.
These early studies utilized a specific reaction model.
We have developed a time-dependent wave
packet method for fully quantum calculations using a three-body model
\cite{Yabana95, Yabana97, Yabana03, Yabana04, Nakatsukasa04}. 
Our results have disagreed with the fusion enhancement.
In fact, the calculation has even suggested a slight suppression of the fusion 
probability because of the presence of the halo neutron.
However, recent coupled-channel calculations using
a similar three-body model, again, indicate
a substantial enhancement in the fusion probability at 
sub-barrier energies \cite{Hagino00, Diaz02}.

Experimental results also show a rather confusing status.
In early measurements, an enhancement of the fusion cross section 
at sub-barrier energies was reported for reactions using
a two-neutron-halo nucleus $^6$He \cite{Kolata98, Trotta00}.
Recently, however, Ref. \cite{Raabe04} have reported that
the enhancement previously obtained in Ref. \cite{Trotta00}
is not due to the fusion but to
the transfer-induced fission.
In case of a one-neutron-halo nucleus, $^{11}$Be,
the experimental results did not present an evident conclusion either
\cite{Signorini98}.
A recent comparison between $^{10}$Be and $^{11}$Be fusion cross sections
suggests no enhancement for $^{11}$Be \cite{Signorini04}.
Thus, we may say, at least, that the present experimental status indicates
no evidence for a fusion enhancement of halo nuclei.

In this letter, we report calculations of the fusion
cross sections in the three-body model.
We previously reported fusion probabilities of halo nuclei 
mostly for the case of total angular momentum $J=0$ which 
corresponds to the head-on collision 
\cite{Yabana03, Yabana04, Nakatsukasa04}. 
We now include a full range of impact parameter, for $0\leq J\leq 30$,
then, for the first time, present results of fusion cross sections
calculated with the time-dependent wave-packet method.
Calculations are performed for fusion cross sections of
$^{11}$Be - $^{209}$Bi and $^6$He - $^{238}$U.
Results are compared with experimental data \cite{Raabe04, Signorini04}. 

In our three-body model,
the projectile is described as a weakly bound two-body system of 
the core and the halo neutron(s).
The projectile ($P$), which is composed of the core ($C$) plus 
neutron(s) ($n$), and the target ($T$) constitute the three-body model. 
For $^{11}$Be, we assume a
$^{10}$Be$+n$ structure in which the halo neutron occupies a $2s$ 
orbital in the projectile ground state. For $^6$He, a di-neutron model 
of $\alpha + 2n$ is assumed. 
The time-dependent Schr\"odinger equation for the three-body model 
is given as
\begin{eqnarray}
&&i\hbar \frac{\partial}{\partial t} \Psi({\bf R},{\bf r},t)
= 
\left\{ -\frac{\hbar^2}{2\mu} \nabla_{\bf R}^2
    -\frac{\hbar^2}{2m}   \nabla_{\bf r}^2 \right.
\nonumber\\
&&   \left. +V_{nC}(r) +V_{CT}(R_{CT}) +V_{nT}(r_{nT}) \right\}
\Psi({\bf R},{\bf r},t),
\label{3BTDS}
\end{eqnarray}
where we denote the relative $n$-$C$ coordinate as ${\bf r}$ and the
relative $P$-$T$ coordinate as ${\bf R}$. The reduced masses of
$n$-$C$ and $P$-$T$ motions are $m$ and $\mu$, respectively.

The Hamiltonian in Eq.~(\ref{3BTDS}) includes interaction potentials
among constituents. The $n$-$C$ potential, $V_{nC}(r)$, is a real 
potential. This potential should produce a halo structure of the projectile.
The core-target potential, $V_{CT}(R_{CT})$, 
consists of the nuclear and Coulomb potentials. The nuclear potential 
is complex and its short-ranged imaginary part is inside the Coulomb barrier 
between the core and the target nuclei. 
The $n$-$T$ potential, $V_{nT}(r_{nT})$, is taken to be real.
We treat the constituents as spin-less point particles, and 
ignore the non-central forces.
The nuclear potentials of $V_{nC}$, $V_{CT}$, and $V_{nT}$ are
taken to be of a Woods-Saxon shape. 
For $^{11}$Be - $^{209}$Bi reaction,
we take $V_0=-53$ MeV, $r_0=1.3$ fm, $a=0.7$ fm for $V_{nC}$,
$V_0=-65$ MeV, $r_0=1.16$ fm, $a=0.8$ fm for the real part of $V_{CT}$,
$V_0=-60$ MeV, $r_0=0.6$ fm, $a=0.4$ fm for the imaginary part of $V_{CT}$,
and
$V_0=-44.019$ MeV, $r_0=1.27$ fm, $a=0.67$ fm for $V_{nT}$.
For $^6$He - $^{238}$U reaction, we adopt
$V_0=-69.305$ MeV, $r_0=0.667$ fm, $a=0.65$ fm ($V_{nC}$),
$V_0=-95$ MeV, $r_0=1.1101$ fm, $a=0.5834$ fm (real $V_{CT}$),
$V_0=-10$ MeV, $r_0=0.6$ fm, $a=0.7$ fm (imaginary $V_{CT}$), and
$V_0=-93.4$ MeV, $r_0=0.986$ fm, $a=1.184$ fm ($V_{nT}$).
The $V_{nC}$ provides the $2s$ bound orbital at $-0.6$ MeV for $^{11}$Be,
and $-0.97$ MeV for $^6$He, which are close to the experimental separation
energies of one and two neutrons, respectively.
The adopted parameter set for $V_{nC}$ is close to that used in the studies 
of the $\alpha$+$^6$He and $p$+$^6$He scatterings 
by Rusek {\it et al.} \cite{6heopt}. 
This parameter set gives a large value of 
$B(E2;g.s.\rightarrow2^+)$ for $^6$He as pointed out in 
Ref.~\cite{6heopt}. 

We define the fusion cross section in terms of the loss of flux
because of the imaginary potential in $V_{CT}$. 
The use of the imaginary potential is approximately
equivalent to the incoming boundary condition inside the barrier.
Once the core and the target nuclei are in the range of the
imaginary $V_{CT}$ inside the Coulomb barrier,
the total flux decreases by the absorption,
no matter whether the neutron is captured by the target or not.
Therefore, this includes not only complete but also a part of 
the incomplete fusion in which the charged core ($^{10}$Be for $^{11}$Be 
and $^4$He for $^6$He) and the target fuse while 
the neutron escapes. 

The wave-packet calculation is performed with the partial wave expansion.
Here, we use the partial wave expansion in the body-fixed frame
\cite{Pack74,Imanishi87}.
Although this is equivalent to the one in the space-fixed frame,
the body-fixed frame is computationally superior for states
at finite total angular momentum, $J\neq 0$.
In the body-fixed frame, the channels are specified by the projection
of $J$ on the body-fixed $z$-axis, $\Omega$,
and the magnitude of the relative angular 
momentum conjugate to the angle between $\mathbf{r}_{nC}$ and $\mathbf{R}_{PT}$
which we denote as $l$. 
The coupling between different $\Omega$ channels is present
only for $\Delta \Omega=\pm 1$, given by the Coriolis term.
This sparse coupling term makes computations
much easier in the body-fixed channel.
Another advantage in the body-fixed frame is a natural 
truncation scheme set by the maximum value of $\Omega$. 

We adopt the model space of the radial coordinates,
$0<R<50$ fm and $0<r<60$ fm, and
discretize them with steps $(\Delta R, \Delta r)=(0.2,1.0)$ fm 
$(\Delta R, \Delta r)=(0.3,0.6)$ fm for $^{11}$Be and 
and $^6$He, respectively. In the case of $^{11}$Be, we also performed 
a calculation with a finer mesh $\Delta r=0.6$ fm and 
confirmed that the fusion probabilities coincide with the 
present results in a good accuracy. 
The maximum $n$-$C$ relative angular momentum is taken
as $l_{\rm max}=70$ to obtain converged results. If the $n$-$T$ potential,
$V_{nT}$
is switched off, one can use a much smaller cutoff, typically $l_{\rm max}=4$.
The discrete-variable representation is used
for the second differential operator \cite{Colbert92}. 
The time evolution of the wave packet is calculated employing the fourth order 
Taylor expansion method. 

The initial wave packet in the incident channel is
a spatially localized Gaussian wave packet multiplied with an incoming wave
for the P-T relative motion while the $n$-$C$ wave function is a
bound orbital corresponding to the halo structure.
The average P-T distance of the wave packet is set as 22 fm and the
average incident energy approximately equal to the barrier top energy. 
The time evolution is calculated until the wave packet consists only of 
outgoing waves and the average P-T distance reaches 30 fm.

A single wave packet solution contains information for 
a certain range of energy. We extract the fusion probability for 
a fixed incident energy by the energy projection procedure \cite{Yabana97}. 

Since the ground state of $^{11}$Be is known 
to be well described by a single neutron halo model, 
our calculation for $^{11}$Be is expected to be realistic.
On the other hand, a di-neutron model for $^6$He may be too crude.
In order to include two-neutron correlation,
we need a three-body $\alpha+n+n$ treatment for $^6$He.
Then, this requires a four-body description for the reaction,
which is beyond the present computational ability. 
We consider the present di-neutron treatment may provide a qualitative 
description for $^6$He. 

\begin{figure}
\includegraphics[width=50mm]{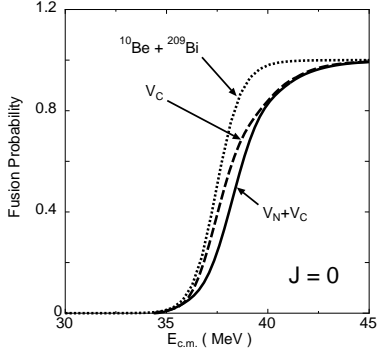}
\caption{\label{fig:11BeJ0}
Fusion probability of $^{11}$Be - $^{209}$Bi at $J=0$.
Calculation with $V_{nT}$ (solid curve) is compared
with that without $V_{nT}$ potential (dashed), and two-body
calculation without a halo neutron (dotted).
}
\end{figure}

\begin{figure}
\includegraphics[width=50mm]{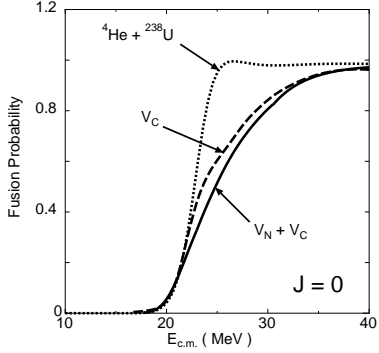}
\caption{\label{fig:6HeJ0}
The same as Fig.~\ref{fig:11BeJ0} but for
$^{6}$He - $^{238}$U.
}
\end{figure}

Now, let us show the calculated results.
In Figs.~{\ref{fig:11BeJ0}} and {\ref{fig:6HeJ0}}, we show
calculated fusion probabilities at the head-on collision ($J=0$)
as a function of the incident energy.
In Fig.~{\ref{fig:11BeJ0}}, the fusion probability of
$^{11}$Be - $^{209}$Bi (solid curve) is compared with that of
$^{10}$Be - $^{209}$Bi (dotted).
The latter is a result of a simple 
two-body calculation with the $V_{CT}$ potential. We also show,
by a dashed line, a
fusion probability of $^{11}$Be calculated without the neutron-target 
potential, $V_{nT}$.
Figure~\ref{fig:11BeJ0} indicates that the fusion probability 
is slightly suppressed by the presence of the halo neutron. The suppression
is observed in both calculations with and without $V_{nT}$. 
This fact suggests that the suppression originates from the core-target 
Coulomb field which induces the dissociation of the halo neutron.
The fusion probability of $^6$He - $^{238}$U system 
also shows a behavior similar to $^{11}$Be
(Fig.~{\ref{fig:6HeJ0}})
The suppression of the fusion probability is seen to be even
stronger than that of $^{11}$Be. This may be due to
a difference of the effective charges of the halo neutron: 
$(1/11)(4e)$ for $^{11}$Be and $(2/6)(2e)$ for $^6$He. 
The Coulomb dissociation effect is more significant for
$^6$He.
There is only a minor difference between the solid and dashed curves 
for both cases, $^{11}$Be and $^6$He.
We have confirmed that the results do not depend on choice of $V_{nT}$.
Therefore, we conclude that reaction mechanism associated with $V_{nT}$
does not play a major role in the fusion process of halo nuclei.
It may be rather surprising that
a long-ranged attractive potential of a halo nucleus, which is expected
in a folding picture of $V_{CT}$ and $V_{nT}$,
does not contribute much to the fusion probability.
This means that the reaction dynamics are very different from
those naturally expected by a simple picture.

In order to understand why,
we should look in detail at the time evolution of the wave packet.
Then, we find the following picture for the dynamics of halo nuclei
\cite{Yabana03}:
When the core nucleus is decelerated by the target Coulomb field,
the halo neutrons keep their incident velocity,
leaving the core nucleus.
This yields the Coulomb breakup of the projectile.
After the breakup, the fusion takes place
between two charged particles, the core and the target nuclei.
The halo neutrons behave like a spectator.
Based on this
picture, we propose
a mechanism that may explain the reason for the slight suppression
of the fusion probability. The incident energy of the projectile
is initially carried by the core and the halo neutron.
After the Coulomb breakup,
since the dissociated neutrons, as a spectator,
keep their shared energy,
the energy of the core nucleus is smaller than the total incident energy of
the projectile. This practical decrease in the incident energy makes the 
fusion probability suppressed for a neutron-halo nucleus in 
comparison with the nucleus without halo neutrons.
In this spectator picture, 
one may naively expect that effect of the breakup simply leads to
an energy shift in curves of the fusion probability.
However, our calculation indicates that the fusion probability 
above the barrier is suppressed, while the probability below 
the barrier is almost unchanged. Thus the effect
is not that simple and depends on the incident energy.

\begin{figure}
\includegraphics[width=50mm]{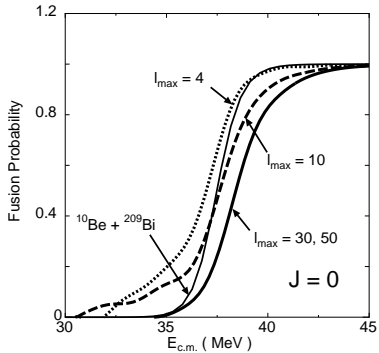}
\caption{\label{fig:11Belmaxdep}
Fusion probability of $^{11}$Be - $^{209}$Bi with different $l_{\rm max}$.
For reference, the fusion probability without
a halo neutron is plotted by a thin-solid curve.
}
\end{figure}

In Refs.\cite{Hagino00,Diaz02}, 
a strongly enhanced sub-barrier fusion probability was reported for halo
nuclei by solving a similar three-body problem. This apparently disagrees with
our results which indicate the suppressed fusion probability.
We found that this discrepancy originates from a slow convergence 
of the calculated result with respect to $l_{\rm max}$ \cite{Yabana03}. 
The treatment of the imaginary potential is also different : we 
include it between the core and the target, while it is in the 
coordinate of the center of mass of the whole projectile 
relative to the target in Ref.~\cite{Hagino00,Diaz02}. 
We examined both treatments of the imaginary potential and found that the 
results are identical. 
In the coupled-discretized continuum-channels (CDCC) approach 
which is adopted in Refs.\cite{Hagino00,Diaz02}, 
a small value of $l_{\rm max}$ is applied for the $n$-$C$ relative angular 
momentum, typically $l_{\rm max}=4$. 
In Fig.~{\ref{fig:11Belmaxdep}}, we show the fusion probability 
for different cutoff values of $l_{\rm max}$.
If we adopt a small $l_{\rm max}$ value,
a strong enhancement of the sub-barrier fusion probability is obtained
in our three-body calculations.
This is consistent to results of the CDCC calculation \cite{Hagino00,Diaz02}.
However, increasing $l_{\rm max}$, the fusion probability 
gradually decreases, and eventually
becomes smaller than the values without a halo neutron (dashed line).
The result of $l_{\rm max}=30$ cannot be distinguished from that of 
$l_{\rm max}=50$. 
To get a converged result, therefore, one needs to adopt $l_{\rm max}\geq 30$. 
We confirmed that this slow convergence with respect to $l_{\rm max}$
is due to the neutron-target potential, $V_{nT}$.
If we ignore $V_{nT}$ and includes only the Coulomb breakup induced 
by $V_{CT}$,
the convergence is much faster, usually $l_{\rm max}=4$ is enough.

Extending the calculations to finite $J$ up to $J=30$,
we achieve the total fusion cross section.
In addition to $l_{\rm max}$, we must also specify a maximum value of
$\Omega$ which we denote as $\Omega_{\rm max}$.
We compare fusion probabilities with $\Omega_{\rm max}=0,5,$ and $10$
for a fixed $J$, and find that the results with $\Omega_{\rm max}=5$ 
and 10 are identical.
Furthermore, a small difference between results of
$\Omega_{\rm max}=0$ and $\Omega_{\rm max}=5$ is visible only for
energies well above the barrier, $E>42$ MeV.
Thus, we here adopt the no-Coriolis approximation ($\Omega_{\rm max}=0$)
to calculate the cross sections.

\begin{figure}
\includegraphics[width=50mm]{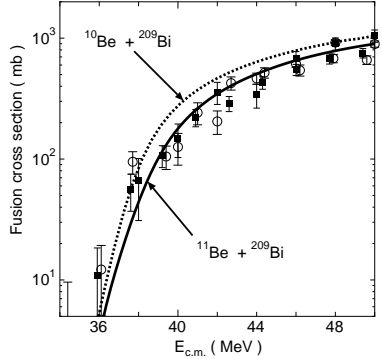}
\caption{\label{fig:11Befull}
Fusion cross section of $^{11}$Be - $^{209}$Bi (solid line) 
and $^{10}$Be - $^{209}$Bi (dotted line). Measured cross
sections \cite{Signorini04} are also shown: $^{11}$Be by 
filled squares and $^{10}$Be by open circles.
}
\end{figure}

We show  calculated fusion cross 
sections of $^{10,11}$Be - $^{209}$Bi in Fig.~{\ref{fig:11Befull}}. 
The $V_{CT}$ for $^{10}$Be - $^{209}$Bi is constructed so that the fusion 
cross section of this system is reproduced by a simple potential 
picture. Adding a halo neutron, the fusion cross section 
decreases slightly for an entire energy region.
This is because the suppression of the fusion probability observed 
in the $J=0$ also appears in finite $J$.

The measurement indicates that the fusion cross sections of $^{11}$Be
do not differ from those of $^{10}$Be within the experimental error.
At this stage, it is difficult to judge whether or not our 
calculated results of slight suppression in fusion cross section
conflict with the measurements.
In our calculation, a part of the halo neutron proceeds to the 
forward direction keeping the incident velocity after fusion. 
Therefore, the observation of these forward neutrons in the incomplete
fusion process may support our picture. In Ref.\cite{Signorini04}, 
the forward neutron emission accompanying the fusion is discussed.

\begin{figure}
\includegraphics[width=50mm]{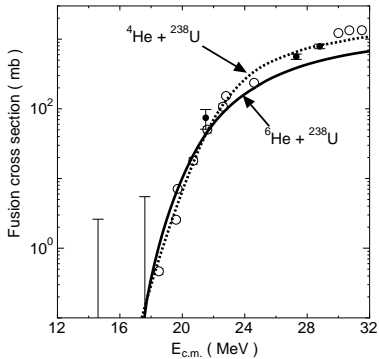}
\caption{\label{fig:4Hepotsc}
The same as Fig.~\ref{fig:11Befull} but for $^{6}$He - $^{238}$U.
Experimental data are taken from Ref. \cite{Raabe04},
shown by open circles ($^4$He) and by filled circles ($^6$He).
Two data points with large error bars at $E_{\rm c.m.}\leq 18$ MeV 
indicate the data for $^6$He. 
}
\end{figure}

We next consider the fusion cross section of $^6$He - $^{238}$U. 
In Fig.~{\ref{fig:4Hepotsc}}, we show the calculated fusion cross
sections of $^6$He and $^4$He on $^{238}$U.
The experimental data in Ref.\cite{Raabe04} indicate, again, that there is 
no difference between $^6$He and $^4$He fusion cross section within 
a statistical error. Our calculation
indicates that the cross section for $^6$He is suppressed from
the one for $^4$He at energies above the barrier.
The magnitude of suppression in $^6$He is larger
than that in $^{11}$Be at high energies.
On the other hand, at sub-barrier energies,
the calculated fusion cross sections of $^4$He and $^6$He are almost
identical to each other.
We again need to wait for measurements with high statistics to make a definite 
conclusion whether our calculation agrees with the experimental data.

In summary, we calculated the fusion cross section for 
nuclei with neutron halo structure. The reaction is described
as the three-body model, and the three-body Schr\"odinger equation 
is solved exactly with the time-dependent wave-packet method.
We compare calculated cross sections with measurements for
$^{11,10}$Be - $^{209}$Bi and $^{6,4}$He - $^{238}$U collisions. 
Recent measurements of these systems suggest that there is no evidence 
for an enhancement in the fusion cross section of halo nuclei. 
Our calculation indicates that the fusion cross section is even 
suppressed by adding a halo neutron. Further measurements with
increased statistics are desirable to obtain definite conclusion
whether the fusion cross section is increased or suppressed by the
presence of the halo neutron.

This work has been supported by the Grant-in-Aid for Scientific
Research in Japan (Nos. 14540369 and 17540231).
A part of the numerical calculations have been performed
at RCNP, Osaka University and at SIPC, University of Tsukuba.


\end{document}